\documentclass[twocolumn,pra]{revtex4-1}

\usepackage[english]{babel}

\usepackage{graphicx}
\usepackage{amsmath}
\usepackage{amsfonts}
\usepackage{amssymb}
\usepackage{bm,bbm}
\usepackage{color}

\usepackage{hyperref}

\newcommand{\sx}[1]{\sigma^{\rm x}_{#1}}
\newcommand{\sy}[1]{\sigma^{\rm y}_{#1}}
\newcommand{\sz}[1]{\sigma^{\rm z}_{#1}}
\newcommand{\sm}[1]{\sigma^{\rm -}_{#1}}
\newcommand{\spp}[1]{\sigma^{\rm +}_{#1}}
\newcommand{\Zjr}[2]{Z^{[#2]}_{#1}}
\def\ii{{\rm i}}

\newcommand{\tr}{\mathrm{tr}}

\def\tit#1{{\em #1},}
\def\etal#1{ {\em et al.}}

\begin{document}

\title{Superdiffusive magnetization transport in the XX spin chain with non-local dephasing}

\author{Marko \v Znidari\v c}
\affiliation{Department of Physics, Faculty of Mathematics and Physics, University of Ljubljana, Jadranska 19, SI-1000 Ljubljana, Slovenia}

\date{\today}

\begin{abstract}
We study a recently discussed XX spin chain with non-local dephasing~\cite{Ren23} in a steady-state boundary-driven setting, confirming superdiffusive magnetization transport in the thermodynamic limit. The emergence of superdiffusion is rather interesting as the Lindblad operators causing it are a coherent sum of two terms, each of which would separately cause diffusion. One therefore has a quantum phenomenon where a coherent sum of two diffusive terms results in superdiffusion. We also study perturbations of the superdiffusive model, finding that breaking the exact form of dissipators, as well as adding interactions to the XX chain, results in superdiffusion changing into diffusion.
\end{abstract}

\maketitle

\section{Introduction}

Transport is one of the simplest nonequilibrium properties which though is not necessarily easy to address, particularly in one-dimensional quantum lattice models. Namely, in one-dimensional systems one can have an interesting interplay between integrability on one hand, which in itself favors ballistic transport where disturbances spread linearly in time, and chaos on the other hand, where one expects diffusion with its square-root growth of disturbances~\cite{review}. Different transport types can be distinguished by a dynamical scaling exponent $z$ that tells us how fast the size of a disturbance, say a width $\sigma$ of a packet, spreads in time, $\sigma \sim t^{1/z}$.

Specific cases with ballistic $z=1$ as well as diffusive $z=2$ were known for a long time. It was also known that in quadratic systems, i.e. systems that are non-interacting in a single-particle basis, one can have an intermediate superdiffusive transport with $1<z<2$ if one allows for an inhomogeneous Hamiltonian, for instance, a site-dependent potential~\cite{foot0}. An example of such a superdiffusive system is the Fibonacci model~\cite{kohmoto83,ostlund83,hiramotoabe}, or a random dimer model~\cite{dunlap90}. Significant progress has been made in the last decade also for interacting models~\cite{review}, with realization that one can have superdiffusion also in a homogeneous interacting system. This was first observed numerically in the isotropic Heisenberg spin chain at infinite temperature~\cite{super} where $z=3/2$. By now we have a fairly thorough understanding of why and when such an ``interacting'' superdiffusion occurs~\cite{sarang19,dupont20,vir20b,jacopo20,enej21,rahul21,ziga20,dupont21,claeys22}, see also Ref.~\cite{virRev} for a review, with a microscopic framework being provided by a generalized hydrodynamics~\cite{GHD1,GHD2}. It is limited to a zero-magnetization sector in integrable models with a continuous non-Abelian symmetry. Integrability is required in order to have ballistically propagating quasiparticles, while the non-Abelian symmetry ensures appropriate properties of those quasiparticles (scaling of their velocity and magnetization they carry with their size). Intriguing was also observation~\cite{kpz,evers20} of not just the scaling exponent $z=3/2$ but also of the associated Kardar-Parisi-Zhang (KPZ)~\cite{KPZ} scaling functions in a fully coherent (noiseless) quantum system. Superdiffusion in the Heisenberg spin chain has been also observed experimentally~\cite{jepsen20,tennant21,bloch22,john23}.

Very recently a surprisingly simple new way of obtaining superdiffusion in quantum lattice models has been revealed~\cite{Ren23}, namely, rather than using symmetry one can use a multi-site dephasing dissipation to induce superdiffusion in an otherwise free fermionic model (equivalent to the XX spin chain). The fact that the dephasing dissipators acts on multiple sites is crucial; for local dephasing one instead gets diffusion~\cite{jstat10}. Superdiffusion comes due to the dephasing strength being zero at some momentum, resulting in a diverging scattering length for those ballistic plane-wave quasiparticles of the XX chain, causing the dynamical scaling exponent $z=3/2$. One can also get other values of $z$~\cite{foot2} if one has a higher order zero in the momentum-space dephasing strength, or if the free-fermion dispersion relation has a zero in the velocity. The mechanism of this newly discovered superdiffusion is different than in previously mentioned interacting integrable models as well as in free inhomogeneous systems; the model is translationally invariant and the phenomenon is not limited to one spatial dimension. Ref.~\cite{Ren23} presented theoretical arguments explaining superdiffusion and verified its prediction by a direct numerical simulation of time evolution of a fully polarized domain wall. The largest size $L=256$ was not large enough to really be in the asymptotic regime of long times, however a hydrodynamic approximation with a Wigner function that though can be simulated in the asymptotic regime did agree with the exact numerics. 

In the present paper we use a boundary-driven Lindblad setting that allows us to (i) probe much larger systems upto $L=6000$, thereby confirming asymptotic superdiffusion in an exact lattice model, (ii) probe the role of weak interactions and weak breaking of dephasing dissipation, both resulting in diffusion, and (iii) verify that the superdiffusion in question is a genuine bulk thermodynamic property and is e.g. not particular to a specific initial state (a fully polarized domain wall can be a non-generic initial state in some situations, like e.g., in the XXZ spin chain, with a non-generic transport type being specific to that state).

\section{Nonequilibrium steady-state setting}

We will use spin language rather than fermions~\cite{Ren23} and study a chain of spin $1/2$ particles with the bulk described by the XX spin chain, written in terms of Pauli operators ($\sx{j},\sy{j},\sz{j}$ and $\sigma^{\pm}_j=(\sx{j}\pm \ii \sy{j})/2$) as  
\begin{equation}
  H=\sum_{j=1}^{L-1} \sx{j}\sx{j+1}+\sy{j}\sy{j+1}.
  \label{eq:XX}
\end{equation}
In fermionic language it describes a system of $L$ spinless non-interacting fermions. On top of the Hamiltonian part we also have a bulk dissipation, such that the evolution equation of the density operator $\rho(t)$ is the Lindblad master equation~\cite{Lindblad1,Lindblad2},
\begin{equation}
  \frac{{\rm d}\rho}{{\rm d}t}=\ii[\rho,H]+\gamma \sum_{j=2}^{L-1}{\cal L}_j^{\rm (deph)}(\rho)+{\cal L}^{\rm (bath)}(\rho).
\label{eq:Lin}
\end{equation}
There are two dissipative parts. The one with ${\cal L}^{(\rm deph)}_j$ of strength $\gamma$ (set to $\gamma=1$) describes bulk non-local dephasing and is the term responsible for interesting superdiffusive transport. The bath part ${\cal L}^{\rm (bath)}$ will act only on the boundary and is there solely to efficiently probe transport properties.

The dephasing superoperator ${\cal L}^{(\rm deph)}_j$ will act on few sites surrounding the site $j$, in our case on $3$ neighboring sites $j-1$, $j$ and $j+1$, and is described by a single Lindblad operator $L_j$ of form $L_j=l_j^\dagger l_j$,
\begin{equation}
  {\cal L}^{\rm (deph)}_j(\rho)= 2L_j \rho L_j^\dagger-\rho L_j^\dagger L_j- L_j^\dagger L_j \rho,\quad L_j=l^\dagger_j l_j.
  \label{eq:Ldeph}
\end{equation}
We will use different forms of $l_j$, resulting in either superdiffusion or diffusion. Just as an example, taking
\begin{equation}
  l_j=\frac{1}{\sqrt{2}}(\sm{j-1}+\Zjr{j-1}{2}\sm{j+1}),
  \label{eq:I}
\end{equation}
where $Z_j^{[r]}$ is a product of $\sz{k}$ on $r$ consecutive sites, starting with the $j$-th,
\begin{equation}
  \Zjr{j}{r}=\prod_{k=j}^{j+r-1}\sz{k},
  \label{eq:Z}
\end{equation}
will result in superdiffusion. Such dissipation is called a non-local dephasing in analogy with the standard (local) dephasing, where one takes $L_j=\spp{j}\sm{j}=(\mathbbm{1}+\sz{j})/2$, as it can be thought of as a dephasing acting on quasiparticles delocalized over few sites~\cite{Ren23}. It is instructive to write out the Lindblad operator $L_j=l_j^\dagger l_j$; for the above choice (\ref{eq:I}) we get
\begin{equation}
  L_j=\frac{2+\sz{j-1}+\sz{j+1}}{4}+\frac{\spp{j-1}\Zjr{j-1}{2}\sm{j+1}-\sm{j-1}\Zjr{j-1}{2}\spp{j+1}}{2}.
  \label{eq:Lsum}
\end{equation}
We can see that for real spins, living on sites $j$, the Lindblad operator is a coherent sum of dephasing and of next-nearest-neighbor hopping. Each of these terms individually is expected to lead to diffusion, for dephasing see Ref.~\cite{jstat10}, for hopping Ref.~\cite{viktor11}, but both together, as we shall see, cause superdiffusion. Interestingly, at first sight a benign looking phase term $\Zjr{j-1}{2}=\sz{j-1}\sz{j}$ is absolutely crucial -- leaving it out in $l_j$ (\ref{eq:I}) will lead to diffusion. Also worth noting is that the Lindblad operator (\ref{eq:Lsum}) is quadratic and Hermitian in terms of fermionic operators using Jordan-Wigner transformation (this will be important for efficient numerics).

To study transport we couple the first and the last spin to magnetization baths described phenomenologically by the following 4 Lindblad operators,
\begin{eqnarray}
  {\cal L}^{\rm (bath)}(\rho)= \sum_{k=1}^4 2L'_k \rho L'^\dagger_{k}-\rho L'^\dagger_k L'_k- L'^\dagger_k L'_k \rho \nonumber\\
 L'_1=\sqrt{\Gamma(1+\mu)}\,\spp{1},\quad L'_2= \sqrt{\Gamma(1-\mu)}\, \sm{1} \nonumber \\
L'_3 =  \sqrt{\Gamma(1-\mu)}\,\spp{L},\quad L'_4= \sqrt{\Gamma(1+\mu)}\, \sm{L}.
\end{eqnarray}
This standard setup~\cite{dario21} can be thought of as an infinite-temperature magnetization driving. The coupling strength $\Gamma$ is set to $\Gamma=1$, while the driving parameter $\mu$ determines magnetization that the bath is trying to impose on the two boundary spins ($\mu$ on $\sz{1}$ and $-\mu$ on $\sz{L}$). We use small $\mu=0.1$ throughout the paper, meaning that we are in a linear response regime where all observables relevant for magnetization transport are proportional to $\mu$.

Lindblad equation (\ref{eq:Lin}) has a single steady-state solution $\rho_\infty$. For $\mu=0$, i.e., no magnetization bias, the steady-state is a trivial infinite-temperature state $\rho_\infty \sim \mathbbm{1}$, for nonzero $\mu$ though it is a true nonequilibrium steady state (NESS) with nontrivial magnetization profile and a site-independent NESS magnetization current $J$.
\begin{figure}[ht!]
  \centerline{\includegraphics[width=2.8in]{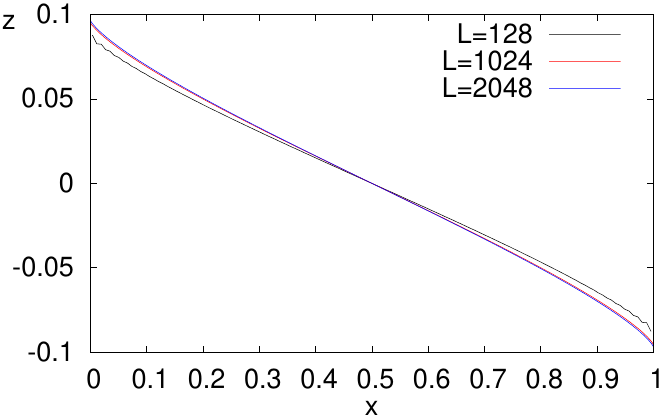}}
  \caption{NESS magnetization profile for non-local dephasing described by eq.(\ref{eq:I}), and $\gamma=\Gamma=1$, $\mu=0.1$. $x$ is a scaled coordinate along the chain. In the thermodynamic limit magnetization smoothly varies from $+\mu$ at the left edge to $-\mu$ at the right edge.}
\label{fig:profil}
\end{figure}
An example of a NESS magnetization profile $\tr{(\rho_\infty \sz{j})}$ for dissipation Eq. (\ref{eq:I}) is shown in Fig.~\ref{fig:profil}. We can see that the profile is not linear, as one would expect for an ordinary diffusion, suggesting superdiffusion.

The type of transport is most easily inferred from the scaling of the NESS current $J$ with system size $L$, keeping driving $\mu$ constant~\cite{review}. Namely, the current will in general have a power law dependence,
\begin{equation}
  J \sim \frac{1}{L^{z-1}},
  \label{eq:z}
\end{equation}
with a dynamical scaling exponent $z$ characterizing transport type. For ballistic transport one has $z=1$, an example is the XX chain without dephasing~\cite{JPA10}, diffusion is characterize by $z=2$, an example being the XX chain with local dephasing~\cite{jstat10}, while $1<z<2$ indicates superdiffusion.

\section{Current scaling}

Let us study the scaling of $J$ with $L$ more in detail. First, because we have a nontrivial dephasing in the bulk, the local current operator is not simply equal to the standard $j_k=2(\sx{k}\sy{k+1}-\sy{k}\sx{k+1})$. Writing the continuity equation for the expectation value $z_j=\tr(\rho_\infty \sz{j})$,
\begin{equation}
  \frac{{\rm d}z_k}{{\rm d}t}=J_{k-1}-J_k,
\end{equation}
defines the local current operator $J_k$, which has in the bulk an additional term due to nonzero $\langle \sz{k},\sum_j {\cal L}^{(\rm deph)}_j(\rho) \rangle$. For instance, for the $l_j$ in Eq.(\ref{eq:I}) we get
\begin{equation}
  J_k=j_k+\frac{\gamma}{2}(\sz{k-1}+\sz{k}-\sz{k+1}-\sz{k+2}).
  \label{eq:JI}
\end{equation}
Due to a 3-site action of ${\cal L}^{(\rm deph)}_j$ the additional term involves 4 sites surrounding the bond $(k)-(k+1)$ across which the current $J_k$ flows. Note that in all cases studied the total magnetization is conserved, $\sum_j {\cal L}^{(\rm deph)}_j(\sum_k \sz{k})=0$.

Crucial for the correct assessment of transport is being able to obtain results for sufficiently large system sizes $L$. If that is not the case one is in danger of making incorrect conclusions~\cite{comment21}. We will use two different numerical methods to obtain NESS $\rho_\infty$, and in turn the NESS current $J=\tr{(J_k\rho_\infty)}$. One is time-evolved-block decimation (TEBD) method~\cite{Schollwock,vidal04}, where the expansion coefficients of $\rho$ in the Pauli basis are written in terms of a product of matrices -- a so-called matrix product operator ansatz. Time evolution by Lindblad equation is then split into small Trotter-Suzuki time steps so that the elementary operation involves two nearest-neighbor spins. Because the dephasing in our case acts on three consecutive sites we write the chain of $L$ spins as a ladder of $L/2$ rungs, so that all operations are indeed nearest-neighbor ones but acting on rungs instead of spins. The price one has to pay is that the local operator space dimension is $4^2$ instead of $4$. Details of our TEBD implementation for the Lindblad equation can be found in Ref.~\cite{njp10}. The method works for any Hamiltonian, not just for the non-interacting XX chain, with the efficiency boiling down to the size of matrices required for a given numerical precision.

The second method works when the equations for all 2-point observables (2-point in the fermionic language) form a closed set. That is, instead of having to solve a system of size $4^L$, one has to deal with a system of $L^2$ linear equations. The method can be applied to Lindblad operators that are Hermitian and quadratic in fermionic operators, and only for the XX chain Hamiltonian. Namely, for such a class of systems one has a set of hierarchical equations, first observed for the XX chain with dephasing~\cite{jstat10} and then generalized~\cite{viktor11,bojan}, see also Refs.~\cite{temme12,giedke13,schiro21}. $k$-point observables form a closed set of linear equations with an inhomogeneous term coming from lower orders. Equations can therefore be solved order by order, starting with $2$-point expectation values. In our spin language those 2-point observables are energy-density like ($r \ge 2$),
\begin{equation}
  A_j^{(r)} = \sx{j} \Zjr{j+1}{r-2}\sx{j+r-1} + \sy{j} \Zjr{j+1}{r-2}\sy{j+r-1},
  \label{eq:A}
\end{equation}
while $A^{(1)}_j=-\sz{j}$, and current-like ($r\ge 2$),
\begin{equation}
  B_j^{(r)} = \sx{j} \Zjr{j+1}{r-2}\sy{j+r-1} - \sy{j} \Zjr{j+1}{r-2}\sx{j+r-1}.
  \label{eq:B}
\end{equation}
There are in total $L^2$ such observables. If we put their NESS expectation value in a vector $\mathbf{y}$, we have to solve a system of linear equations
\begin{equation}
  M \mathbf{y}=\mu \mathbf{m},
  \label{eq:M}
\end{equation}
where a sparse matrix $M$ depends on the dephasing strength $\gamma$ and the bath coupling strength $\Gamma$, while a constant source vector $\mathbf{m}$ comes solely due to bath driving. Both our superdiffusive examples, Eq.(\ref{eq:Lsum}) and Eq.(\ref{eq:II}), do posses such hierarchical structure of NESS correlations and so Eq.(\ref{eq:M}) can be used to study large systems. Because of dealing with a finite system, where one has to correctly write equations also at the boundary, the form of $M$ is a bit messy and we give details in Appendix~\ref{app}.

\subsection{Superdiffusion with $z=3/2$}

\begin{figure}[ht!]
  \centerline{\includegraphics[width=3.2in]{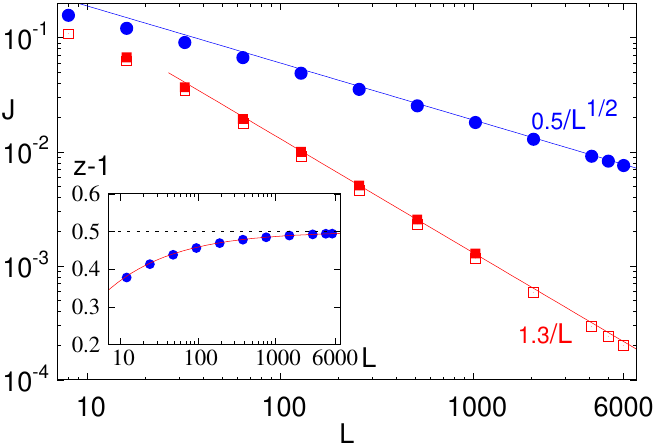}}
  \caption{Scaling of the NESS current $J$ with system's length $L$ for hierarchy-preserving dephasing in Eq.(\ref{eq:I}), blue circles, showing superdiffusion with $z=3/2$. Red squares are for dissipation Eq.(\ref{eq:Ib}) showing diffusion with $z=2$. The inset shows convergence of $z-1$ with system size, with the full curve suggesting a slow $\sim 1/\sqrt{L}$ convergence.}
\label{fig:I}
\end{figure}

Let us start with the dissipator already mentioned in Eq.(\ref{eq:Lsum}), namely $L_j=l_j^\dagger l_j$ with
\begin{equation}
l_j=\frac{1}{\sqrt{2}}\left(\sm{j-1}+\sz{j-1}\sz{j}\sm{j+1}\right).
\end{equation}
For such $l_j$ one has a closed set of $L^2$ linear equations for $2-$point observables (\ref{eq:A},\ref{eq:B}), and we study current in the NESS in systems with upto $L=6000$ spins. Results are shown in Fig.~\ref{fig:I}, where we can see that the dynamical scaling exponent is indeed $z=\frac{3}{2}$. This confirms theoretical prediction for our dissipator in Eq.(\ref{eq:Lsum}) based on Ref.~\cite{Ren23} with numerically exact lattice simulation.
We can also see (the inset) that the convergence of $z$ with $L$ is rather slow, $[\frac{3}{2}-z(L)] \sim 1/\sqrt{L}$. This is in line with a slow convergence with time observed in Ref.~\cite{Ren23} of a hydrodynamic Wigner function approximation for a unitary evolution of a domain wall initial state.

\subsubsection{Breaking superdiffusion}

Considering we are working in spin language it is natural to ask, what would happen if we would use a simpler-looking $l_j$ without the product of two $\sz{k}$ (that are due to the Jordan-Wigner transformation). To that end we take
\begin{equation}
  l_j=\frac{1}{\sqrt{2}}\left( \sm{j-1}+\sm{j+1}\right).
  \label{eq:Ib}
\end{equation}
Current for such $l_j$ stays the same (\ref{eq:JI}). At first sight the difference between Eq.(\ref{eq:I}) and Eq.(\ref{eq:Ib}) is minuscule -- sometimes such phase factors are simply neglected when doing Jordan-Wigner transformations as they are believed not to be important. In our case they are crucial. This new $L_j$ obtained from the operator in Eq.~(\ref{eq:Ib}) is a sum of terms that are quadratic in fermions as well as terms that are quartic (remember, previous Lindblad operator (\ref{eq:Lsum}) had only quadratic terms). The closed hierarchy is broken because one can get 2-point observables from 4-point ones, for instance ${\cal L}^{(\rm deph)}_2(\sx{1}\sx{2}\sz{3})$ ($\sx{1}\sx{2}\sz{3}$ is quartic) will contain also $\sx{2}\sx{3}$ (quadratic).

This means that we can not anymore use the efficiently solvable Eq.(\ref{eq:M}). Rather, in order to get the exact NESS we have to use full TEBD (full red squares in Fig.~\ref{fig:I}). Despite the broken hierarchy we have in addition to the TEBD tried another approximate method. As mentioned, for such $l_j$ (\ref{eq:Ib}) the 2-point observables do not form a closed set anymore; one instead has equations of form $M\mathbf{y}+N\mathbf{w}=\mu \mathbf{m}$, where $\mathbf{w}$ are expectation values of higher point observables. In other words, in order to calculate 2-point functions one needs also higher-point observables (3-point, 4-point,...). The approximation we make is simply dropping all higher point expectations from the above equation, i.e. $\mathbf{w}=0$, and solving resulting equations for 2-point expectation values, see Appendix~\ref{app:Lsup0}. Those results are shown with empty red squares in Fig.~\ref{fig:I}. We can see that the approximation with only 2-point observables works surprisingly well. E.g., at $L=1000$ the difference is about $10\,\%$, but more importantly, the scaling looks to be the same $\sim 1/L$~\cite{foot3}. Therefore, both TEBD and the 2-point approximation show clear diffusion. The conclusion therefore is that as soon as we break a closed hierarchy of correlations for Eq.(\ref{eq:Lsum}) one gets diffusion.

\subsection{Superdiffusion with $z=5/3$}

\begin{figure}[t!]
  \centerline{\includegraphics[width=3.2in]{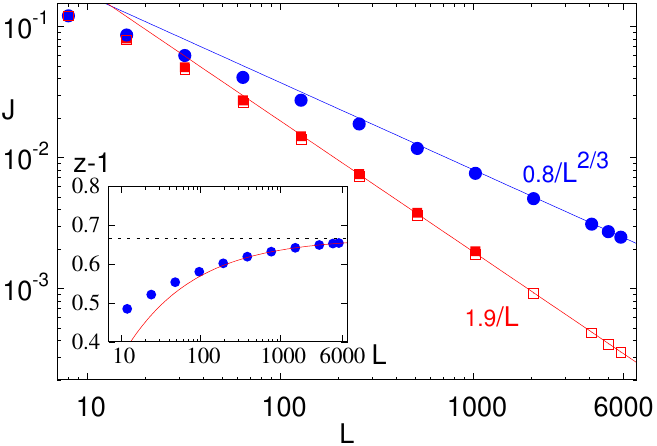}}
  \caption{Scaling of the NESS current $J$ with $L$. Blue circles are for dissipator in Eq.~(\ref{eq:II}) that respects hierarchical structure of correlations, showing superdiffusion with $z=5/3$, while the red squares are for Eq.(\ref{eq:IIb}) without the phase factors and which shows diffusion with $z=2$. The inset shows convergence of $z-1$ with system size, with the full curve suggesting $\sim 1/\sqrt{L}$ asymptotics.}
\label{fig:II}
\end{figure}
It was predicted in Ref.~\cite{Ren23} that the exponent is not always $z=3/2$. It depends on the order of a zero at $k_0$ in the momentum-space dephasing strength, as well as on special points $k_0$ where the velocity of free quasiparticles might be zero. An example of such higher order zero is dephasing dissipators $L_j=l_j^\dagger l_j$ with
\begin{equation}
  l_j=\frac{1}{\sqrt{6}}\left(\sm{j-1}-2\Zjr{j-1}{1}\sm{j}+\Zjr{j-1}{2}\sm{j+1}\right),
  \label{eq:II}
\end{equation}
which is a particular case of a more general
\begin{equation}
  l_j=\frac{1}{\sqrt{2+a^2}}(\sm{j-1}-a\Zjr{j-1}{1}\sm{j}+\Zjr{j-1}{2}\sm{j+1}),
  \label{eq:a}
\end{equation}
studied in Ref.~\cite{Ren23}. For $-2<a<2$ one expects $z=\frac{3}{2}$ while the chosen $a=2$ is marginal with a 2nd order zero and prediction~\cite{Ren23} that the dynamical scaling exponent is $z=5/3$. We again calculate the NESS and the scaling of current, which in this case (\ref{eq:II}) is
\begin{eqnarray}
  \label{eq:JII}  
  J_k=j_k&+&\frac{\gamma}{18}(\sz{k-1}+9\sz{k}-9\sz{k+1}-\sz{k+2})+\\
  &+&\frac{\gamma}{9}(A^{(2)}_{k-1}-A^{(2)}_{k+1})+\frac{\gamma}{9}(A^{(3)}_{k-1}-A^{(3)}_{k}).\nonumber
\end{eqnarray}
The dissipator (\ref{eq:II}) preserves the 2-point expectations and we can study large systems (Appendix \ref{app:Lsup3c}). In Fig.~\ref{fig:II} we show results, demonstrating clear convergence to theoretical prediction. Let us also note that the additional terms in the current expression in Eq.(\ref{eq:JII}) as well as in Eq.(\ref{eq:JI}) are all differences of operators on neighboring sites. Because the steady state expectations are continuous in the spatial index $k$ (see Fig.\ref{fig:profil}) they all scale as $\sim 1/L$, and therefore in the thermodynamic limit for superdiffusion one has $J_k \approx j_k$.

\subsubsection{Breaking superdiffusion}

Next, we check what happens if we remove the phase factors in the above $l_j$, that is, if we take
\begin{equation}
  l_j=\frac{1}{\sqrt{6}}\left( \sm{j-1}-2\sm{j}+\sm{j+1} \right).
  \label{eq:IIb}
\end{equation}
The corresponding current operator is
\begin{eqnarray}
  J_k=j_k&+&\frac{\gamma}{18}(\sz{k-1}+9\sz{k}-9\sz{k+1}-\sz{k+2})+\\
  &+&\frac{\gamma}{9}(A^{(3)}_{k-1}-A^{(3)}_{k}).\nonumber 
\end{eqnarray}
The results are shown in Fig.~\ref{fig:II}, with full red squares for TEBD simulations, and empty red squares using a 2-point correlation function approximation (Appendix \ref{app:Lsup3}), similarly as in the previous subsection for $z=3/2$. Again, we can see that already this subtle change leads to diffusion.

\subsection{Interactions}

\begin{figure}[tb!]
  \centerline{\includegraphics[width=1.65in]{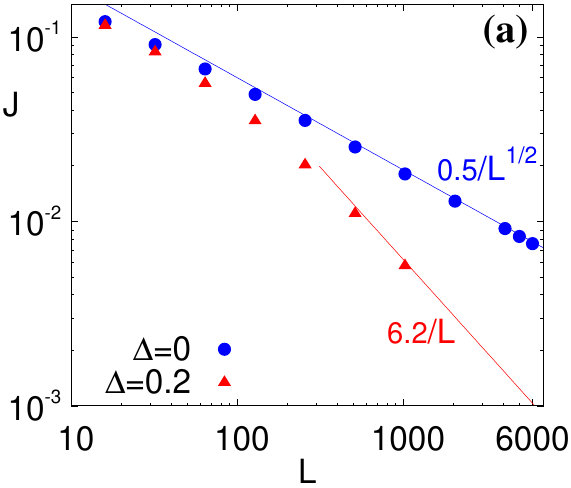}\includegraphics[width=1.65in]{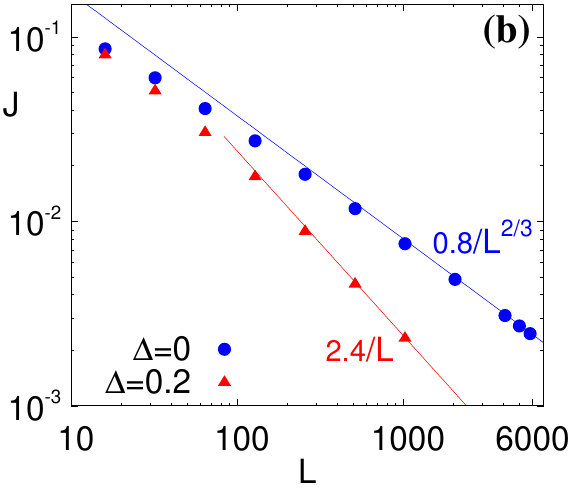}}
  \caption{Interactions cause superdiffusion to go into diffusion. Frame (a) shows dephasing in Eq. (\ref{eq:I}), and (b) dephasing in Eq. (\ref{eq:II}). Blue circles are the non-interacting case, i.e., the same data as in Figs.~\ref{fig:I} and~\ref{fig:II}, while red triangles are for the XXZ chain (\ref{eq:XXZ}) with interaction strength $\Delta=0.2$.}
\label{fig:int}
\end{figure}

We have seen that while one does get superdiffusion for a whole class of dissipators parameterized by $a$ (\ref{eq:a}), superdiffusion goes away if we remove product of $\sz{j}$ in the definition of $l_j$. In this subsection we test what happens in we keep the form of $l_j$ but add interactions to the Hamiltonian. To this end we study the XXZ chain,
\begin{equation}
  H=\sum_{j=1}^{L-1} \sx{j}\sx{j+1}+\sy{j}\sy{j+1}+\Delta \sz{j}\sz{j+1},
  \label{eq:XXZ}
\end{equation}
where $\Delta$ represents interaction. We use TEBD to get the NESS with which we can go upto $L=1024$ with rather modest matrix sizes. As we can see in Fig.~\ref{fig:int} we obtain diffusion for both $z=3/2$ dephasing in Eq. (\ref{eq:I}), and for $z=5/3$ dephasing in Eq. (\ref{eq:II}), already for relatively small interaction $\Delta=0.2$.

While one might jump to a conclusion that this is expected and that superdiffusion is only a property of the XX chain and the specific form of a non-local dephasing, upon reflection things are not that clear. Namely, one can heuristically understand the emergent superdiffusion in the XX chain in the following way~\cite{Ren23}: looking at $l_j$ (\ref{eq:a}) in momentum space, one finds that such non-local $l_j$ results in a momentum dephasing strength that depends on the momentum $k$. This dephasing strength can in particular have a zero at some $k_0$, resulting in a diverging scattering length of free (quasi)particles at that $k_0$. Superdiffusion therefore emerges from a measure zero of non-dephasing ballistic quasiparticles. Following this explanation one could argue that because the XXZ chain is integrable, and as such also harbors ballistic quasiparticles, the same phenomenon should be possible. The important difference compared to the XX chain is that the transformation to quasiparticles is not a simple Fourier transformation and one would have to construct a dephasing that would be zero for those non-plane wave quasiparticles. On the other hand, for the XXZ model one does not have a closed hierarchy of correlations~\cite{foot1}, which seem to be important to get superdiffusion -- breaking that, as we have seen, immediately leads to diffusion, even in the XX model. Therefore more studies are needed to clarify the generality or speciality of the proposed superdiffusion scenario due to non-local dephasing.

\section{Conclusion}

We have demonstrated superdiffusive magnetization transport in the XX spin chain in the presence of non-local dephasing dissipation in the bulk. The effect is simple and interesting: one can view it as the emergence of superdiffusion out of a coherent sum of two diffusive contributions in Lindblad operators. It is different than other known cases of superdiffusion, for instance the one in integrable models with a non-Abelian symmetry, or in inhomogeneous non-interacting systems. We have not touched upon superdiffusive classical systems, however, what we can say is that it seems to be qualitatively different than the superdiffusion observed in the stochastic momentum exchange model~\cite{olla06,mejia09} where the effect changes with dimension.

While a number of questions has been answered, many remain, and some new arose. For instance, during TEBD simulations we have observed that the required size of matrices can be very small. In other words, the operator Schmidt spectrum of the NESS seems to decay quickly. That seems to be the case for both superdiffusive, and to a lesser extent also for diffusive cases studied. One question is can any of the superdiffusive NESSs be written in a matrix product operator form with a low-rank matrices? The fact that the numerical rank is small is perhaps related to two known similar cases of low-rank NESS: for the XX chain and our boundary driving but without dephasing the ballistic NESS requires matrices of size $4$ (independent of $L$)~\cite{JPA10}, while in the presence of the local dephasing the same holds in the leading order in the thermodynamic limit~\cite{pre11}. Related to that, can any of the non-local dephasing cases be exactly solved, for instance along the lines of formal integrability as e.g. in Ref.~\cite{chiara21}.

As discussed, a possibility of superdiffusion under non-local dephasing in other non-free interacting systems remains unclear. Exciting is also an option of having superdiffusion in more than one dimension.

\section*{Acknowledgments}

Support by Grants No.~J1-4385 and No.~P1-0402 from the Slovenian Research Agency is acknowledged.

\appendix

\section{Closed equations for 2-point functions}
\label{app}
When one has a hierarchical structure of equations for our specific bath driving one can write the NESS as
\begin{equation}
 \label{eq:NESS}
 \rho_{\infty} = \cfrac{1}{2^L}[1 + \mu({\cal A} + {\cal B})] + \mathcal{O}(\mu^2)
\end{equation}
where observables that are linear in $\mu$ are in fermionic language a 2-fermion observables, which in spin language read,
\begin{eqnarray}
  {\cal A} &=& \sum_{r=1}^L\sum_{j=1}^{L+1-r}a_j^{(r)}A_j^{(r)},\\
 A_j^{(r+1)} &=& \sx{j} Z_{j+1}^{[r-1]}\sx{j+r} + \sy{j} Z_{j+1}^{[r-1]}\sy{j+r},\quad \hbox{for} r>0 \nonumber
\end{eqnarray}
while for $r=0$ we have $A^{(1)}_j=-\sigma_j^{z}$. The ${\cal B}$ term is on the other hand
\begin{eqnarray}
  {\cal B} &=& \sum_{r=2}^L\sum_{j=1}^{L+1-r}b_j^{(r)}B_j^{(r)} \nonumber\\
  B_j^{(r+1)} &=& \sigma_j^x Z_{j+1}^{[r-1]}\sigma_{j+r}^{y} - \sigma_j^y Z_{j+1}^{[r-1]}\sigma_{j+r}^{x} \nonumber
\end{eqnarray}
Note that the form in Eq.(\ref{eq:NESS}) is exact, and not just an expansion in $\mu$~\cite{jstat10,viktor11,MarkoHorvat,bojan}. Namely, higher order terms in $\mu$ are all orthogonal to $A's$ and $B's$. All unknown expansion coefficients $a_j$ and $b_j$ can be put compactly into a hermitian correlation matrix
\begin{equation}
C_{j,k}=a_j^{(k-j+1)} + \textrm{i}\, b_j^{(k-j+1)},\quad k>j,
\end{equation}
diagonal is $C_{j,j} = a_j^{(1)}$, while $C_{j,k} = C_{k,j}^{*}$ for $j>k$.

Therefore, finding expectation value of any 2-point observable in NESS involves solving a set of linear equations for unknown $C$. Following Ref.~\cite{MarkoHorvat} the steady state equations can be written in a matrix form as,
\begin{equation}
  2\ii (JC-CJ)+2(D C +C D)+\gamma \tilde C-2\mu P=0,
  \label{eq:lyap}
\end{equation}
with the only nonzero matrix elements of $L\times L$ matrices $J,D,P$ being $J_{k,k+1}=J_{k+1,k}=-1$, $P_{1,1}=-2\Gamma, P_{L,L}=2\Gamma$, $D_{1,1}=D_{L,L}=\Gamma$. The first term in eq.(\ref{eq:lyap}) is due to Hamiltonian, the 2nd and 4th due to boundary driving, and the 3rd term due to dephasing. First three terms are linear in the correlation matrix, while the 4th term is a constant driving term and so the eq.(\ref{eq:lyap}) represents a set of $n^2$ linear equations for unknown $C$.

Matrix $\tilde{C}$ accounts for dephasing, depends linearly on $C$, and is more complicated due to a 3-site action of ${\cal L}^{(\rm deph)}_j$, as well as boundary effects. Its form can be obtained by evaluating action of ${\cal L}^{(\rm deph)}_j$ on the ansatz (\ref{eq:NESS}). In the following sections we will just list the result for each of the four dissipators used.

\subsection{Dephasing with $z=3/2$}
\label{all:Lsup0c}

Take dephasing in Eq.(\ref{eq:I}) that results in $z=3/2$ superdiffusion, i.e. for
\begin{equation}
  l_j=\frac{1}{\sqrt{2}}(\sm{j-1}+Z^{[2]}_{j-1}\sm{j+1}).
\end{equation}
Off diagonal elements $\tilde{C}_{i,i+r}$ with $|r|\ge 3$ can be expressed as
\begin{equation}
  \tilde C=LC+CL,\quad L=\frac{1}{2}
    \begin{bmatrix}
      1 &  & 1\cr
       & 1 &  & \ddots \cr
      1 &  & 2 &  & \cr
      & \ddots & & \ddots & \cr
      & & & & 2 & & 1\cr
      & & & & & 1 & \cr
      & & & & 1 &  & 1 \cr
    \end{bmatrix}.
    \label{eq:LI}
\end{equation}
Matrices of this form (almost Toeplitz matrices with boundary effect) will appear in all cases, so we will use a shorter way of defining them by simply listing only nonzero elements. For instance, the above matrix (\ref{eq:LI}) is specified by $L_{k,k+2}=L_{k+2,k}=\frac{1}{2}$, $L_{k,k}=(\frac{1}{2},\frac{1}{2},1,\ldots,1,\frac{1}{2},\frac{1}{2})$.

Elements on $r$-diagonals $\tilde{C}_{i,i+r}$ with $r=0,1,2$ must on the other hand be written separately (dephasing acting on $k$ neighboring sites changes the form of $\tilde{C}_{j,j+r}$ for $|r|<k$). To shorten notation, let us put all the elements on the $r$-diagonal, i.e. $\tilde{C}_{i,i+r}$, into a vector $\mathbf{\tilde{c}}_r$, and likewise for $\mathbf{c}_r=\{C_{j,j+r}\}$. For instance, $\mathbf{c}_2=(C_{1,3},C_{2,4},\ldots,C_{L-2,L})$, with an additional convention that when a square matrix of size larger than $L-r$ acts on $\mathbf{c}_r$ we add a sufficient number of zeros at the end of vector $\mathbf{c}_r$.

Then the main diagonal, i.e. $r=0$, of $\tilde{C}$ is given by $\mathbf{\tilde{c}}_0=R^{(0)} \mathbf{c}_0$, where the $L$-dimensional matrix $R^{(0)}$ has nonzero elements $R^{(0)}_{j,j+2}=R^{(0)}_{j+2,j}=-\frac{1}{2}$ and $R^{(0)}_{j,j}=(\frac{1}{2},\frac{1}{2},1,\ldots,1,\frac{1}{2},\frac{1}{2})$.

The $1$-diagonal is given by $\mathbf{\tilde{c}}_1=R^{(13)} \mathbf{c}_3+R^{(11)} \mathbf{c}_1+R^{(11c)} \mathbf{c}^*_1$, where the $(L-1)$-dimensional matrix $R^{(13)}$ has nonzero elements $R^{(13)}_{j,j}=R^{(13)}_{k+2,k}=\frac{1}{2}$, while nonzero elements of $R^{(11)}$ are $R^{(11)}_{j,j}=(1,\frac{3}{2},2,\ldots,2,\frac{3}{2},1)$, and $R^{(11c)}_{j,j+1}=R^{(11c)}_{j+1,j}=\frac{1}{2}$.

The $2$-diagonal, $r=2$, is given by $\mathbf{\tilde{c}}_2=R^{(24)} \mathbf{c}_4+\mathbf{c}_2+R^{(22c)} \mathbf{c}^*_2$, with $(L-2)$-dimensional $R^{(24)}_{j,j}=R^{(24)}_{k+2,k}=\frac{1}{2}$, and $R^{(22c)}_{j,j}=(-\frac{1}{2},-\frac{1}{2},0,\ldots,0,-\frac{1}{2},-\frac{1}{2})$.

Matrix elements of $\tilde{C}$ below the diagonal are determined from those above by hermiticity, $\tilde{C}^\dagger=\tilde{C}$.

\subsection{Dephasing with $z=5/3$}
\label{app:Lsup3c}
Here $l_j$ is given by Eq.(\ref{eq:II}), that is
\begin{equation}
  l_j=\frac{1}{\sqrt{6}}(\sm{j-1}-2\sz{j-1}\sm{j}+ Z^{[2]}_{j-1}\sm{j+1}).
\end{equation}
Matrix elements of $\tilde{C}$ on all $r$-diagonals with $|r|\ge 3$ can be again expressed as $\tilde C=LC+CL$, with nonzero elements of $L$ being $L_{k,k}=\frac{1}{6}(1,5,6,\ldots,6,5,1)$, $L_{k,k+1}=L_{k+1,k}=-\frac{1}{3}(1,2,\ldots,2,1)$, $L_{k,k+2}=L_{k+2,k}=\frac{1}{6}$.

The main diagonal is instead equal to $\mathbf{\tilde{c}}_0=R^{(00)} \mathbf{c}_0+R^{(01)} (\mathbf{c}_1+\mathbf{c}^*_1)/2+R^{(02)} (\mathbf{c}_2+\mathbf{c}^*_2)/2$, with nonzero elements of $R^{(00)}$ being $R^{(00)}_{j,j}=-\frac{1}{18}(5,13,18,\ldots,18,13,5)$, $R^{(00)}_{j+1,j}=R^{(00)}_{j,j+1}=\frac{2}{9}(1,2,\ldots,2,1)$, and $R^{(00)}_{j+2,j}=R^{(00)}_{j,j+2}=\frac{1}{18}$. Nonzero elements of $L$-dimensional $R^{(01)}$ are $R^{(01)}_{j,j}=\frac{2}{9}(2,1,\ldots,1,-1,1)$, $R^{(01)}_{j,j+1}=-\frac{2}{9}$, $R^{(01)}_{j+1,j}=\frac{2}{9}(-1,1,\ldots,1,2)$, and $R^{(01)}_{j+2,j}=-\frac{2}{9}$. Nonzero elements of $L$-dimensional $R^{(02)}$ are $R^{(02)}_{j,j}=-\frac{2}{9}$, $R^{(02)}_{j+1,j}=\frac{4}{9}$, and $R^{(02)}_{j+2,j}=-\frac{2}{9}$.

The $1$-diagonal is $\mathbf{\tilde{c}}_1=R^{(10)} \mathbf{c}_0+R^{(11r)} \mathbf{c}_1+R^{(11c)}\mathbf{c}^*_1+R^{(12r)} \mathbf{c}_2+R^{(12c)} \mathbf{c}^*_2+R^{(13)} \mathbf{c}_3$, with nonzero $R^{(13)}_{j,j}=R^{(13)}_{k+2,k}=\frac{1}{6}$ of a $(L-1)$-dimensional $R^{(13)}$. Nonzero elements of $(L-1)$-dimensional $R^{(12r)}$ are $R^{(12r)}_{j,j}=-\frac{1}{9}(5,\ldots,5,2,5)$ and $R^{(12r)}_{j+1,j}=-\frac{1}{9}(2,5,\ldots,5)$, while nonzero elements of $R^{(12c)}$ are $R^{(12c)}_{j,j}=R^{(12c)}_{j+1,j}=\frac{1}{9}$. Nonzero elements of $(L-1)$-dimensional $R^{(11r)}$ are $R^{(11r)}_{j,j}=\frac{1}{18}(14,25,28,\ldots,28,25,14)$, $R^{(11r)}_{j,j+1}=R^{(11r)}_{j+1,j}=\frac{2}{9}$, while of $R^{(11c)}$ are $R^{(11c)}_{j,j}=-\frac{2}{9}(1,2,\ldots,2,1)$ and $R^{(11c)}_{j,j+1}=R^{(11c)}_{j+1,j}=-\frac{1}{18}$. Finally, nonzero elements of $L$-dimensional $R^{(10)}$ are $R^{(10)}_{j,j}=-\frac{1}{9}(2,1,\ldots,1)$, $R^{(10)}_{j,j}=-\frac{1}{9}(2,1,\ldots,1,-1,1)$, $R^{(10)}_{j,j+1}=-\frac{1}{9}(-1,1,\ldots,1,2)$ and $R^{(10)}_{j,j+2}=R^{(10)}_{j+1,j}=\frac{1}{9}$ (in $R^{(10)} \mathbf{c}_0$ only the first $(L-1)$ components go into $\tilde{c}_1$).

The $2$-diagonal is $\mathbf{\tilde{c}}_2=R^{(20)} \mathbf{c}_0+R^{(21r)} \mathbf{c}_1+R^{(21c)}\mathbf{c}^*_1+R^{(22r)} \mathbf{c}_2-\frac{1}{18}\mathbf{c}^*_2+R^{(23)} \mathbf{c}_3+R^{(24)} \mathbf{c}_4$. Nonzero elements of $(L-2)$-dimensional $R^{(22r)}$ are $R^{(22r)}_{j,j}=\frac{2}{9}(5,8,\ldots,8,5)$. Nonzero elements of $(L-1)$-dimensional $R^{(21r)}$ are $R^{(21r)}_{j,j}=-\frac{1}{9}(5,5,\ldots,5,2,5)$ and $R^{(21r)}_{j,j+1}=-\frac{1}{9}(2,5,\ldots,5)$, while $R^{(21c)}_{j,j}=R^{(21c)}_{k,k+1}=\frac{1}{9}$ (only the first $(L-2)$ components of $R^{(21r)} \mathbf{c}_1+R^{(21c)}\mathbf{c}^*_1$ matter). Nonzero elements of $(L-2)$-dimensional $R^{(24)}$ are $R^{(24)}_{j,j}=R^{(24)}_{k+2,k}=\frac{1}{6}$, while of the same-sized $R^{(23)}$ are $R^{(23)}_{j,j}=-\frac{1}{3}(2,2,\ldots,2,1,2)$ and $R^{(23)}_{j+1,j}=-\frac{1}{3}(1,2,\ldots,2)$. Finally, nonzero elements of $L$-dimensional $R^{(20)}$ are $R^{(20)}_{j,j}=-\frac{2}{9}$ and $R^{(20)}_{j,j+1}=R^{(20)}_{j+1,j}=\frac{1}{9}$ (middle $(L-2)$ components of $R^{(20)} \mathbf{c}_0$ is what goes into $\mathbf{\tilde{c}}_2$).

\subsection{Perturbed $z=3/2$ dephasing}
\label{app:Lsup0}
The operators $l_j$ are (\ref{eq:Ib})
\begin{equation}
  l_j=\frac{1}{\sqrt{2}}(\sm{j-1}+\sm{j+1}).
\end{equation}
Remember that in this case one does not have a closed set of equations for 2-point observables. Nevertheless, as explained, dropping higher order expectation values in equations for 2-point functions gives a good approximation (red squares in Fig.\ref{fig:I}). Elements of $\tilde{C}_{j,j+r}$ with $|r|\ge 3$ are equal to corresponding matrix elements of $LC+CL$ with diagonal $L_{k,k}=\frac{1}{2}(1,2,3,\ldots, 3,2,1)$.

The main diagonal is instead $\mathbf{\tilde{c}}_0=R^{(0)} \mathbf{c}_0$, with nonzero $R^{(0)}_{j,j}=-\frac{1}{2}(1,1,2,\ldots,2,1,1)$, $R^{(0)}_{j,j+2}=R^{(0)}_{j+2,j}=\frac{1}{2}$. The $1$-diagonal is $\mathbf{\tilde{c}}_1=R^{(1)} \mathbf{c}_1$ with $R^{(1)}_{j,j}=\frac{1}{2}(2,3,4,\ldots,4,3,2)$. The $2$-diagonal is given by $\mathbf{\tilde{c}}_2=R^{(2)} \mathbf{c}_2-\frac{1}{2}\mathbf{c}^*_2$, with $R^{(2)}_{j,j}=\frac{1}{2}(3,4,\ldots,4,3)$.

\subsection{Perturbed $z=5/3$ dephasing}
\label{app:Lsup3}
In this case the operators $l_j$ are (\ref{eq:IIb})
\begin{equation}
  l_j=\frac{1}{\sqrt{6}}(\sm{j-1}-2\sm{j}+\sm{j+1}).
\end{equation}
Neglecting higher order correlations, elements of $\tilde{C}_{j,j+r}$ with $|r|\ge 3$ are equal to the matrix elements of $LC+CL$, with $L_{k,k}=\frac{1}{18}(3,16,19,\ldots,19,16,3)$, $L_{k,k+1}=L_{k+1,k}=\frac{1}{18}(5,10,\ldots,10,5)$, $L_{k,k+2}=L_{k+2,k}=\frac{1}{9}$.

Elements $\tilde{C}_{j,j+r}$ with $r=0,1,2$ must instead be given separately. The main diagonal is $\mathbf{\tilde{c}}_0=R^{(0)} \mathbf{c}_0+R^{(02)}\mathbf{c}'_2$, where $\mathbf{c}'_2=(a^{(1)}_1,(\mathbf{c}_2+\mathbf{c}^*_2)/2,a^{(1)}_{L-1})$, and nonzero elements $R^{(0)}_{j,j}=-\frac{1}{18}(5,13,18,\ldots,18,13,5)$, $R^{(0)}_{j,j+1}=R^{(0)}_{j+1,j}=\frac{2}{9}(1,2,\ldots,2,1)$, $R^{(0)}_{j,j+2}=R^{(0)}_{j+2,j}=\frac{1}{18}$, as well as $R^{(02)}_{j,j}=\frac{1}{9}(-3,4,\ldots,4,-3)$, $R^{(02)}_{j,j+1}=-\frac{1}{9}(2,\ldots,2,-3)$ and $R^{(02)}_{j+1,j}=-\frac{1}{9}(-3,2,\ldots,2)$.

The $1$-diagonal is $\mathbf{\tilde{c}}_1=R^{(12)} \mathbf{c}_2+R^{(13)}\mathbf{c}_3+R^{(11r)}\mathbf{c}_1+R^{(11c)}\mathbf{c}^*_1$. Nonzero elements of $(L-1)$-dimensional $R^{(13)}$ are $R^{(13)}_{j,j}=R^{(13)}_{k+2,k}=\frac{1}{9}$. Nonzero elements of $(L-1)$-dimensional $R^{(12)}$ are $R^{(12)}_{j,j}=\frac{1}{9}(4,\ldots,4,\frac{3}{2},4)$ and $R^{(12)}_{j+1,j}=\frac{1}{9}(\frac{3}{2},4,\ldots,4)$. Nonzero elements of $(L-1)$-dimensional $R^{(11r)}$ are $R^{(11r)}_{j,j+1}=R^{(11r)}_{j+1,j}=-\frac{2}{9}$ and $R^{(11r)}_{j,j}=\frac{1}{18}(14,25,28,\ldots,28,25,14)$, while for $R^{(11c)}$ we have $R^{(11c)}_{j,j}=-\frac{2}{9}(1,2,\ldots,2,1)$ and $R^{(11c)}_{j,j+1}=R^{(11c)}_{j+1,j}=\frac{1}{9}$.

The $2$-diagonal is $\mathbf{\tilde{c}}_2=R^{(20)} \mathbf{c}_0+R^{(21)} \mathbf{c}_1+R^{(22r)} \mathbf{c}_2-\frac{1}{18}\mathbf{c}^*_2+R^{(23)} \mathbf{c}_3+R^{(24)} \mathbf{c}_4$. Nonzero elements of $(L-2)$-dimensional $R^{(24)}$ are $R^{(24)}_{j,j}=R^{(24)}_{j+2,j}=\frac{1}{9}$. Nonzero elements of $(L-2)$-dimensional $R^{(23)}$ are $R^{(23)}_{j,j}=\frac{5}{9}(1,\ldots,1,\frac{1}{2},1)$ and $R^{(23)}_{j+1,j}=\frac{5}{9}(\frac{1}{2},1,\ldots,1)$. Nonzero elements of $(L-2)$-dimensional $R^{(22r)}$ are $R^{(22r)}_{j,j}=(\frac{7}{6},\frac{17}{9},\ldots,\frac{17}{9},\frac{7}{6})$. Nonzero elements of $R^{(21)}$ are $R^{(21)}_{j,j}=\frac{4}{9}(1,\ldots,1,\frac{3}{8},1)$ and  $R^{(21)}_{j,j+1}=\frac{4}{9}(\frac{3}{8},1,\ldots,1)$. Finally, nonzero elements of $R^{(20)}$ are $R^{(20)}_{j,j}=-\frac{2}{9}$ and $R^{(20)}_{j,j+1}=R^{(20)}_{j+1,j}=\frac{1}{9}$.


\begin{thebibliography}{}
\bibitem{foot0} In inhomogeneous quadratic systems one can in fact construct an arbitrary transport type, see e.g. Ref.~\cite{mantica}.
\bibitem{foot2} Predicted exponents do not necessarily follow Fibonacci sequence~\cite{slava15}.
\bibitem{foot3} One reason why higher $r$-point correlations can be neglected could be due to them scaling as $\mu^r$ (such a scaling is observed in the XX chain with local dephasing~\cite{pre11}) and using small $\mu=0.1$.
\bibitem{foot1} It is worth mentioning that the closeness of correlations is not always limited solely to noninteracting models, see Ref.~\cite{bojan}.
\end{thebibliography}
\end{document}